\documentclass[prb,aps,twocolumn,superscriptaddress,floatfix]{revtex4}
\pdfoutput=1

\usepackage{color,graphicx,bm,amsmath}

\newcommand{\ec}{\ensuremath{E_{\mathrm{cap}}}}
\newcommand{\mc}{\ensuremath{\mu_c}}

\newcommand{\dsas}{\ensuremath{\Delta}}

\newcommand{\mum}{\ensuremath{\mu\mathrm{m}}}

\newcommand{\ei}{\ensuremath{\varepsilon_i}}

\newcommand{\dd}{\ensuremath{d_d}}
\DeclareMathOperator{\sgn}{sgn}

\begin{document}

\title{Vortex states of a disordered quantum Hall bilayer}

\author{P. R. Eastham}\affiliation{Blackett Laboratory, Imperial
College London, London SW7 2AZ, United Kingdom}

\author{N. R. Cooper}\affiliation{Cavendish Laboratory, 
University of Cambridge, Cambridge CB3 0HE, United Kingdom}

\author{D. K. K. Lee}\affiliation{Blackett Laboratory, Imperial
College London, London SW7 2AZ, United Kingdom}

\date{\today}

\begin{abstract} 
  We present and solve a model for the vortex configuration of a
  disordered quantum Hall bilayer in the limit of strong and smooth
  disorder. We argue that there is a characteristic disorder strength
  below which vortices will be rare, and above which they proliferate.
  We predict that this can be observed tuning the electron density in
  a given sample. The ground state in the strong-disorder regime can
  be understood as an emulsion of vortex-antivortex crystals. Its
  signatures include a suppression of the spatial decay of counterflow
  currents. We find an increase of at least an order of magnitude in
  the length scale for this decay compared to a clean system.  This
  provides a possible explanation of the apparent absence of leakage
  of counterflow currents through interlayer tunneling, even in
  experiments performed deep in the coherent phase where enhanced
  interlayer tunneling is observed.
\end{abstract}


\maketitle

\section{Introduction}

There has been much recent progress in the search for quantum
condensed phases of quasiparticles in solids, such as Bose-Einstein
condensates of excitons, polaritons, and magnons. A very interesting
example\cite{eisenstein04,murphy_many-body_1994,lay_anomalous_1994}
occurs for electron bilayers in the quantum Hall regime. When the two
layers are close, and have individual filling factors $\nu=1/2$, the
Coulomb interactions produce a ground state in which electrons in one
layer are correlated with holes in the other. The wavefunction of this
state is that of a Bose-Einstein condensate of interlayer excitons,
and it exhibits behaviors reminiscent of superfluidity and the
Josephson effects: a small counterflow
resistivity,\cite{tutuc04,kellog04} which can be understood as
excitonic superfluidity, and a zero-bias tunneling
anomaly,\cite{spielman00,tiemann08} which can be interpreted as a
Josephson effect. However, the analogy is incomplete, because neither
the counterflow resistivity nor the width of the tunneling anomaly\
\cite{eisenstein04-ssc} appears to vanish at finite temperatures.

Many theoretical works have suggested that these deviations from
conventional superfluid behaviors are connected to the presence of
vortices. In a quantum Hall system physical and topological charges are
related, so that random electric fields, created by the dopants, could
induce vortices. The hypothesis that this leads to a disordered vortex
state has been used\cite{fogler01,sterngirvinma,radzihovsky01} to
explain features such as the width of the tunneling anomaly and the
region of negative differential conductance. More recently, Fertig and
collaborators have developed a strong-disorder model, in which the
dissipation reflects the dynamics of a vortex
liquid.\cite{fertig,roostaei08} Despite these potential consequences,
however, there have been few attempts to predict the vortex
configuration in a bilayer. For weak, layer-antisymmetric disorder the
appropriate model is a gauge glass,\cite{sterngirvin,fertig}
suggesting vortex liquids, glasses, or conventional superfluid states
are
possibilities.\cite{sterngirvin,radzihovsky01,fertig,fisher89,girvin95}
This is supported by exact diagonalization\cite{sheng03} of small
systems with white noise disorder.

The aim of this paper is to predict the vortex configuration of a
quantum Hall bilayer, for the case of strong, long-range disorder, as
is experimentally relevant for high mobility modulation doped
samples. We argue that for a fixed disorder potential there is a
characteristic value of the magnetic length, above which vortices
proliferate. We find that this proliferation corresponds to the
formation of an emulsion of vortex-antivortex crystals. Our theory
should be testable, since we estimate that the proliferation occurs in
an experimentally accessible regime. Furthermore, we argue that the
proliferation causes a dramatic suppression of the decay of
counterflow currents. We find a new length scale for this decay which
is one to two orders of magnitude larger than the corresponding
length scale in the clean system. This provides a possible explanation
of a long-standing puzzle of the persistence of counterflow
currents\cite{kellog04} across an entire sample, in a regime where
enhanced interlayer tunneling conductance is observed.  Such behavior
is quantitatively confirmed in recent experiments which show an area
scaling for tunneling currents\cite{eisenstein-area} up to the scale
of $100 \mum$. More generally, our work suggests that the quantum Hall
bilayer could be used to study a disordered form of the
``supersolid''\cite{leeteitel,vavcrystals} that has previously
attracted attention in superfluids, superconductors, and a clean
bilayer model.\cite{tupitsyn96}

The remainder of this paper is structured as follows. In
Sec.~\ref{sec:model} we develop a model for the vortex
configuration of the bilayer, and identify the parameters which
control the vortex density. In Sec.~\ref{sec:ground-states} we
present numerical results for the ground state of the model, and
compare these with a mean-field theory of an emulsion. In
Sec.~\ref{sec:discussion} we analyze the decay of counterflow
currents in the ground state, suggest some further consequences of the
emulsion, and discuss the role of antisymmetric disorder. Finally,
Sec.~\ref{sec:conclusions} summarizes our conclusions.

\section{Model}
\label{sec:model}

\begin{figure}[h]
\includegraphics[width=0.4\textwidth]{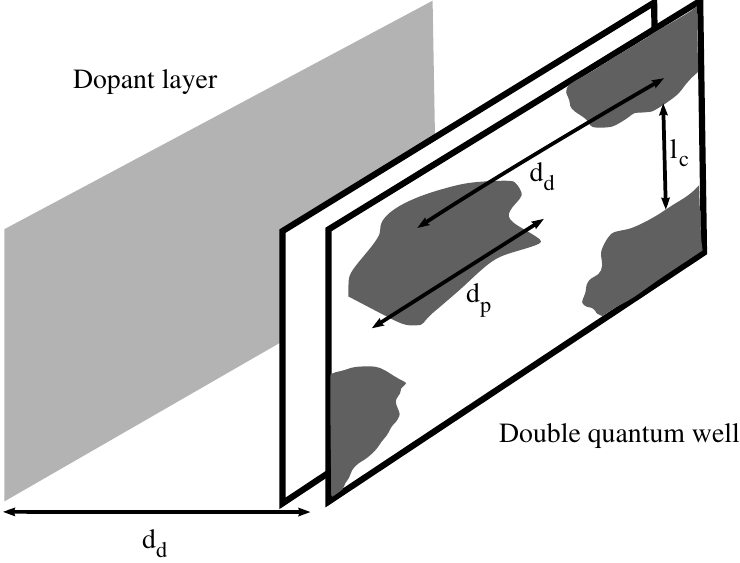}
\caption{Schematic of a disordered quantum Hall bilayer with
  compressible puddles of electron liquid (dark-shaded areas) of size
  $d_p$ surrounded by channels of incompressible excitonic superfluid
  of size $l_c$. For smooth disorder $d_p\sim l_c\sim d_d$ and the
  depicted length scales are larger than the magnetic length $l_0$. In
  the limit of very strong disorder, $l_c$ can become small and
  comparable to $l_0$. \label{fig:schematic}}
\end{figure}

We begin by developing a model for the vortex configuration, which we
solve both numerically and in a mean-field approximation. Our starting
point is the ``coherence network'' picture,\cite{fertig} in which
the bilayer consists of compressible puddles of electron liquid,
separated by channels of the incompressible counterflow superfluid (see
Fig.~\ref{fig:schematic}).  This is appropriate for the strong, smooth
disorder produced by dopants, which destroys the superfluid over a
significant fraction of the sample.\cite{efros,cooper93} We
initially consider only layer-symmetric disorder, since the distance
to the dopants is much larger than the interlayer separation. We focus
on the simplest case of a balanced bilayer, where the filling fraction
in each layer is $\nu=1/2$, and initially also neglect the small
interlayer tunneling.

This picture leads us to postulate the Hamiltonian
\begin{equation}
H=\frac{1}{2} \sum_{ij} (Q_i-{\bar q}_i)E_{ij}(Q_j-\bar q_j) 
+ \frac{1}{2}\sum_{i\neq j} v_i G_{ij}v_j.
\label{eq:basich}
\end{equation} 
The first term is the electrostatic energy of an inhomogeneous charge
distribution, written in terms of the charge $Q_i$ on the $i^{th}$
compressible puddle, and the inverse capacitance matrix of the puddles
$E_{ij}=C_{ij}^{-1}$. The potential due to the dopants is contained in
the continuous-valued shifts $\bar q_i$ which would be the optimum
charges on the puddles in classical electrostatics. This Coulomb term
was not considered by previous work on the coherence network. We will
see that it is the competition between Coulomb energy and superfluid
stiffness that controls the proliferation of vortices in the system.

The second term in Eq.~ \ref{eq:basich} models the energy of the
channels. The condensate is characterized by a local phase
$\theta(r)$, describing the interlayer phase coherence. Since this
phase can wind by integer multiples of $2\pi$ around each puddle we
associate vorticities $v_i$ with the puddles. The superfluid energy in
the channels is $H_{\mathrm{sf}}=\int (\rho_s/2) |\nabla
\theta(\mathbf{r})|^2 d^2\mathbf{r} $, with stiffness\
\cite{moon95,wen_neutral_1992} $\rho_s\sim l_0^{-1}$. As usual,
$H_{\mathrm{sf}}$ leads to a vortex-vortex interaction $G_{ij}\sim
-\log r_{ij}$, and a constraint $\sum_i v_i=0$.

The topological defects of the condensate are merons,\cite{moon95}
which are vortices whose core corresponds to an unpaired electron in
one layer.  The meron charge is $q=(e/2) \sigma v$, where $v$ is the
vorticity, and $\sigma=\pm 1$ denotes the layer index of the core.
Because of this relationship the two terms in Eq.~\ref{eq:basich} are
coupled, and the charge disorder can drive vorticity in the channels.
For those puddles with $|\bar q/e-[\bar q/e]|>1/4$ the electrostatics
favors a half-electron charge, which is allowed if the vorticity
around the puddle $v$ is odd. This costs a superfluid energy
proportional to $v$.  Therefore, a puddle will have $|v|=1$ if this
incurs a superfluid energy cost smaller than the electrostatic energy
gain.

These considerations allow us to identify the parameter controlling
the vortex density in the percolating channels. For a wide range of
parameters both the channel width $l_c$ and puddle size $d_p$ will be
on the order of the distance to the dopant layers, $\dd \approx 200
\mathrm{nm}$.\cite{eisensteinpc} (See Fig.~\ref{fig:schematic} for
illustration of these length scales.) The largest contribution to the
electrostatic energy is the Coulomb interactions within each
puddle. Thus we estimate the electrostatic energy gain of a vortex as
$\ec\sim (1/2) (e/2)^2/C$, where $C\sim d_p\sim d_d$ is the
self-capacitance of the puddle.  We estimate the superfluid energy
cost as the prefactor of the vortex energy, which is generally
$E_s\sim 2\pi \rho_s\sim l_0^{-1}$. Thus the vortex density is
controlled by the ratio
\begin{equation}
\frac{\ec}{E_s}\sim \frac{l_0}{d_d}.
\label{eq:tuningratio}
\end{equation} 
Since $d_d$ is fixed by the
sample, we expect the vortex density to vary with the magnetic length.

In the limit of very strong disorder\cite{fertig} $l_c$ becomes of the
order the magnetic length $l_0\approx 20 \mathrm{nm}$, while $d_p$
remains of order $\dd$. The vortex energy in this regime is $E_s\sim
2\pi \rho_s (l_c/d_d)$, with the factor $l_c/d_d$ accounting for the
fraction of the area occupied by the superfluid (up to numerical
factors depending on the shapes of the puddles). Thus in the strong
disorder limit the vortex density becomes independent of $l_0$,
$\ec/E_s\sim 1$. We estimate this numerical parameter by modeling the
puddles as disks of radius $\dd\approx 200\mathrm{nm}$, and taking
$\rho_s$ from the mean-field theory\cite{moon95} at zero interlayer
separation. This gives $\ec\approx E_s \approx 1\mathrm{K}$.

Since our estimates of $\ec$ and $E_s$ in the strong disorder limit
are comparable, it may be possible to vary the density of vortices in
experiments. Decreasing $l_0$ should take the system further from the
(not unrealistic\cite{fertig,efros,cooper93}) strong disorder limit,
and so could lead to a reduction in the vortex density. More
generally, reducing the vortex density requires a decrease in the
capacitative energies, perhaps by placing gates on both sides of the
sample as close as possible to the wells, or increasing the superfluid
energy, perhaps in samples with smaller interlayer separation and
larger tunneling.

To predict the vortex density and configuration of the bilayer, we now
derive and solve a Hamiltonian for the vorticity. For simplicity we
consider $E_{ij}=2\delta_{ij} \ec$. The off-diagonal terms will not
qualitatively affect the results, because the off-site Coulomb
interactions have a much shorter range than the vortex interactions
$G_{ij}$. The diagonal elements are approximately constant, because
they are controlled mainly by the characteristic puddle size. The main
source of randomness is in the offset charges $\bar q_i$.

Taking $e/2$ as our unit of charge, we write the total charge on each
puddle as $Q_i=q_i^M+\sigma_i v_i$, where $\sigma=\pm 1$, $v=0, \pm
1$, and $q_i^M$ is the meron-free charge. In the ground state $q^M_i$
is the nearest even integer to $\bar q_i$. Thus the electrostatic
energy of a vortex $v_i$ on site $i$ is
\begin{equation}\label{eq:esenergy} E_i= E_{\mathrm{cap}} [v_i^2 +
  2 \sigma_i v_i (q_i^M-\bar q_i)].\end{equation} This is the only
energy contribution which depends on the layer index of the core
$\sigma_i$, and $E_i$ can be minimized by setting $\sigma_i=-\sgn[v_i
(q_i^M-\bar q_i)]$. The distribution of $\bar q$ is broad on the scale
of the charge quantization, because the puddles contain many
electrons, so that $q_i^M-\bar q_i$ is a uniformly distributed random
variable between $\pm 1$. Thus, we see that the electrostatic energy takes the form
$H=\sum_i \ei v_i^2$, where $\ei$ varies randomly from site to site,
with distribution $P(\ei)$. In the approximation that
$E_{\mathrm{cap}}$ is the same for all puddles, $\ei$ is uniformly
distributed between $\pm E_{\mathrm{cap}}$. Note that in reality there
will be some variation in $E_{\mathrm{cap}}$ from puddle to puddle,
and the sharp edges in $P(\ei)$ at $\pm E_{\mathrm{cap}}$ will be
smoothed out.

Combining the electrostatic and superfluid energies, we thus have an
effective Hamiltonian for the vorticity
\begin{equation} 
    H=\sum_i \ei v_i^2 + \frac{1}{2} \sum_{i\neq j} v_i G_{ij}
    v_j. \label{eq:disordermodel}
  \end{equation} We note that the random field is coupled to the presence
  of vortices, independently of their sign. This differs from gauge
  glass models, where the random field couples directly to the
  vorticity.

\begin{figure}[t]
\includegraphics[width=240pt,height=170pt]{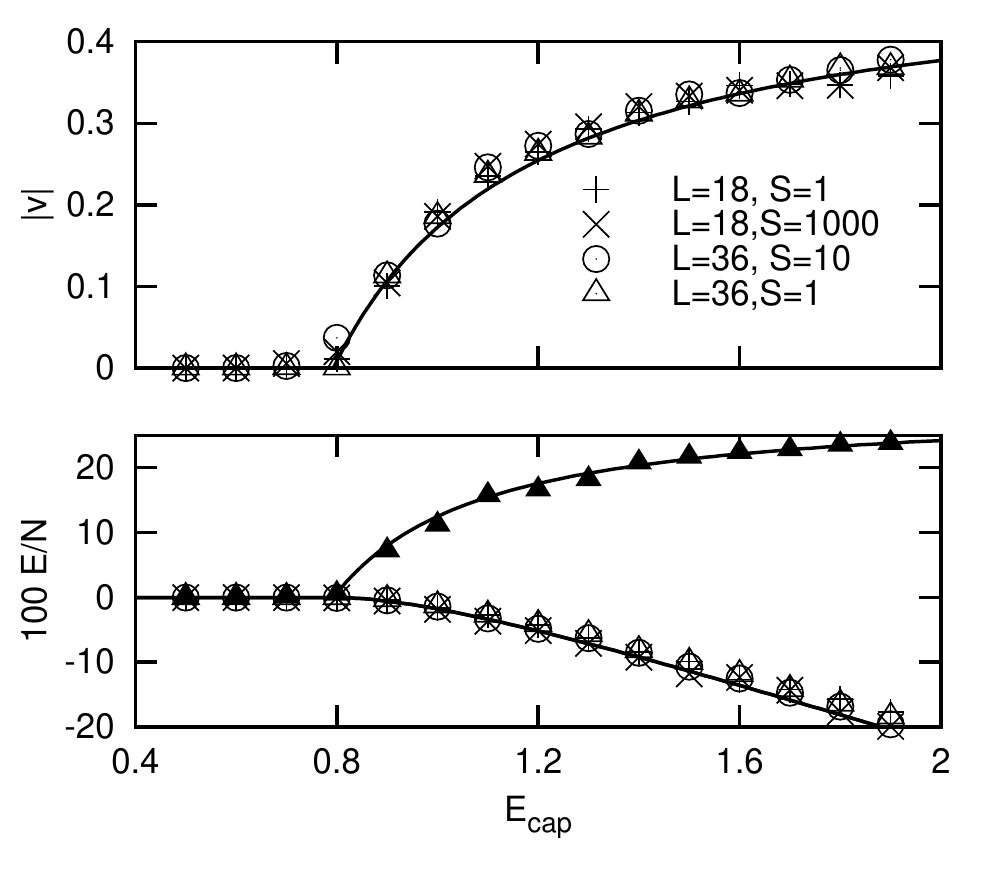}
\caption{Top panel: average ground-state vorticity $|v|$ as a function
  of disorder energy scale $\ec$, obtained by simulated annealing on
  systems of linear size $L$, with $S$ Monte-Carlo sweeps per
  temperature step (see text). Each point is an average over $25$
  disorder realizations. $\ei$ is taken from a uniform distribution of
  width $2\ec$. Bottom panel: corresponding ground state energies
  (crosses and hollow symbols), and interaction energy for $L=18,S=1$
  (solid triangles). Solid curves show the corresponding results of
  the mean-field theory. \label{fig:gsv}}
\end{figure}

\section{Ground states}
\label{sec:ground-states}

\begin{figure}[t]
\includegraphics{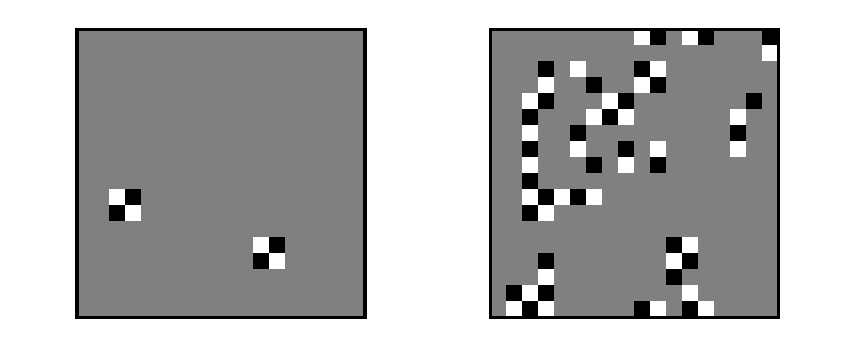}
\caption{Ground states for a typical realization of the disorder at
  strengths $\ec=0.8$ (left) and $1.1$ (right).  Black/white are
  vorticities $\pm 1$, and gray is $0$. $L=18$ and
  $S=1000$.\label{fig:gstates}}\end{figure}

Numerical results for the ground states of Eq.~\ref{eq:disordermodel}
are shown in Figs.\ \ref{fig:gsv} and\ \ref{fig:gstates}. We adopt a
lattice model (as in Ref.\ \onlinecite{fertig}) where the channels are
the edges of a square lattice of side $L$. We take the prefactor of
the vortex energy $E_s$ to be our unit of energy. Thus
$G_{ij}=V(\mathbf{r}_{ij})-V(0)$ is the lattice solution to $\nabla^2
V(\mathbf{r})=-2\pi\delta(0)$, with the singularity
removed.\cite{leeteitel} Ground states were obtained by simulated
annealing, with standard nearest-neighbor Monte Carlo moves. Each
ground state is obtained by recording the lowest energy state obtained
during an anneal, from a temperature of $0.5$ to a temperature of
$0.01$, in steps of $0.01$. At each temperature we perform $S$ sweeps
of $4L^2$ moves. As can be seen in Fig.\ \ref{fig:gsv}, increasing $S$
by a factor of $10^3$ does not significantly change the results, so we
are obtaining good approximations to the ground states. The results in
Fig.\ \ref{fig:gsv} are quenched averages of Monte Carlo data obtained
for different disorder realizations.

From the top panel of Fig.\ \ref{fig:gsv} we see that the ground state
is a uniform superfluid for small $\ec$, while vortices proliferate
above a threshold $\ec^{0}$. The threshold behavior in $|v|$ as a
function of $\ec$ is sharp due to the discontinuity in the on-site
energy distribution $P(\ei)$, and would in reality be rounded due to
the variations of $\ec$ between puddles.

Fig.\ \ref{fig:gstates} shows ground states obtained for a typical
disorder realization at two different strengths. These results show
that the vortex ground states are not completely disordered, and are
strongly suggestive of an emulsion of vortex-antivortex crystals. This
structure appears because the field in Eq.~\ref{eq:disordermodel} does
not dictate the sign of the vorticity. On the square lattice there is
a minimum in the interaction $G_{\mathbf{q}}=\pi/8=\mu_c$ at
wavevector $\mathbf{q}=(\pi,\pi)$, so for a uniform field $\ei<-\mu_c$
the ground state is a vortex-antivortex crystal.\cite{leeteitel} Whereas a random field coupling to $v_i$ (as in a
gauge glass model\cite{fertig,sterngirvin}) competes with this
ordering, the random field coupling to $v_i^2$ does not. It can
therefore straightforwardly induce regions of the crystalline phase.

The vortex density in Fig.\ \ref{fig:gsv} appears to be consistent
with a mean-field theory of an emulsion. To develop such a theory, let us
consider a mixture of two phases occupying fractions $\bar x$ and
$(1-\bar x)$ of the system, with energy densities $\mu_c$ and $0$
respectively.  Without a random field the mean-field energy of such a
mixture is\cite{adkins}
\begin{equation} E=\mu_c \bar x + \kappa \bar x (1-\bar
  x).\label{eq:mfe-hom}\end{equation} $\kappa$ is an interaction
parameter, which corrects for the use of bulk energy densities in the
first term. It will be the only fitting parameter in the theory.

To incorporate the random $\ei$, we interpret Eq.~\ref{eq:mfe-hom} as a
mean-field approximation for the microscopic effective
Hamiltonian
\begin{equation} 
H=\sum_i h x_i + \sum_{\langle ij\rangle} 2J_{ij} x_i x_j, 
\label{eq:pairwiseham}
\end{equation} where $x_i=0$ denotes a site
in the vortex-free phase, and $x_i=1$ one in the vortex crystal
phase. The mean-field approximation is obtained by writing $x_i=\bar x
+ (x_i - \bar x)$, and discarding terms quadratic in the
fluctuations. Demanding that the resulting energy agree with
Eq.~\ref{eq:mfe-hom} allows us to relate $h$ and the average $J_{ij}$ to
$\mu_c$ and $\kappa$. We then incorporate the random field term from
Eq.~\ref{eq:disordermodel}, $\sum_i \ei x_i$, to obtain 
\begin{equation}
H_{\mathrm{mf}}=
\bar x^2 \kappa N+\sum_i (\mc +\ei+\kappa-2\kappa \bar
x)x_i.
\label{eq:mfeffham-inh}
\end{equation}
The mean-field equation is $\bar x=\langle x_i \rangle$, where
$\langle\rangle$ denotes an average in the ground state of
Eq.~\ref{eq:mfeffham-inh}. For the uniform distribution of width
$2\ec$ for $\ei$, we find that, when $\mc +\kappa<\ec$,
\begin{equation} \bar x=\frac{1}{2}\left(1-\frac{\mc
    }{\ec-\kappa}\right),
\label{eq:that-vd}
\end{equation} 
and $\bar x=0$ otherwise. We can also compute the energy, 
\begin{equation}
  \frac{1}{N}\langle H_{\mathrm{mf}}\rangle = \bar x ^2 \kappa + \int_{\mc
    +\kappa-\ec-2\kappa\bar x}^0 \frac{E}{2\ec} dE.
\label{eq:mfen}\end{equation}
The solid lines in Fig.\ \ref{fig:gsv} show the mean-field predictions
of Eqs.~\ref{eq:mfe-hom}, \ref{eq:that-vd}, and \ref{eq:mfen}, with
$\kappa=0.4$ chosen to give the threshold $\ec$ obtained numerically.
As can be seen, this theory, with a single fitting parameter, gives a
good account of the numerical results. Thus the ground state vorticity
of Eq.~\ref{eq:disordermodel} can indeed be understood in terms of the
formation of an emulsion of vortex crystals.

\section{Discussion}
\label{sec:discussion}

The presence of the vortex-crystal emulsion would affect counterflow
and tunneling experiments. Let us consider, in particular, the decay
of a d.c.~counterflow current due to tunneling. Without the vortices,
the superfluid phase $\theta$ is obtained by minimizing the energy
\begin{equation}
H=\int \left[\frac{\rho_s}{2} |\nabla \theta|^2 - \dsas n \cos (\theta)
\right]d^{2}\mathbf{r},
\label{eq:phaseenergy}
\end{equation}
where $\dsas$ is the tunneling strength, and $n=1/(2\pi l_0^2)$ is the
electron density. A small static perturbation to the solution
$\theta=0$, such as a small counterflow current injected at one edge,
decays on the scale set by the Josephson length
\begin{equation}
\lambda_{J}\sim l_0\sqrt{\frac{\rho_{s}}{\dsas}}
\label{eq:josephsonlength}
\end{equation}
estimated\cite{fogler01} as $\sim 5\mu \mathrm{m}$. This means we
should not expect counterflow currents to persist over more than a few
microns due to leakage by interlayer tunneling (in other words, by the
recombination of the interlayer excitons).  This appears inconsistent
with the experimental observation\cite{eisenstein-area} of an area
scaling for the tunneling anomaly, up to length scales of $100 \mum$.

With pinned vortices, we should instead consider the energy associated
with the vorticity-free part of the supercurrents.\cite{fogler01} If
we write the phase field of the vortices as $\theta_0$, we can
separate out the vorticity-free phase field $\phi = \theta-\theta_0$.
For a fixed vortex field $\theta_0$, the ground state of the system is
determined by a random-field XY model for the vorticity-free part of
the system:
\begin{equation}
H_\phi =\int \left[\frac{\rho_s}{2} |\nabla \phi|^2 - 
\dsas n \cos (\phi+\theta_0)
\right]d^{2}\mathbf{r}.
\label{eq:phaseenergy_phi}
\end{equation}
This may be treated using standard techniques.\cite{fukuyama-lee,imryma} In the emulsion, the pinning phase
$\theta_0$ is disordered. We see that, in the limit of a vanishing
correlation length for $\theta_0$, the tunneling field has no effect
because it averages to zero. In our case, the vortex phase field has a
correlation length $\xi \sim d_d \ll \lambda_{J}$, corresponding to
the weak-disorder regime of the random field model. In this regime,
the ground state $\phi$ consists of domains of linear size $L_{\rm
  dom}$, aligned with the average random field across the domain. The
typical tunneling energy in the random field is given by the sum of
random energies in the range $\pm\dsas n\xi^2$ for $(L_{\rm
  dom}/\xi)^2$ correlation areas. This gives a typical energy of
$\dsas(\xi/l_0)^2\sqrt{(L_{\rm dom}/\xi)^2}$.  The cost in phase
stiffness in the domain is of the order of $\rho_s (L_{\rm dom})^0$ in
two dimensions. Balancing these two energies, we find the domain size
\begin{equation}
L_{\rm dom} \sim \lambda_J \left(\frac{\lambda_J}{\xi}\right).
\label{eq:domainsize}
\end{equation}
We estimate that the Josephson length $\lambda_{J}\sim 5\mum$ while
the correlation length $\xi\sim 100\mathrm{nm}$. Therefore, this
domain size $L_{\rm dom}$ is a \emph{new} length scale associated with
the emulsion which could be one to two orders of magnitudes larger
than $\lambda_{J}$ in the clean system. Moreover, we see that static
perturbations to this disordered ground state ($\phi\rightarrow
\phi+\delta\phi$), such as an injected counterflow current, decay over
this new length scale $L_{\rm dom}$.  Allowing for the considerable
uncertainty in $\lambda_J$, this decay length ($\sim 0.3$mm) predicted
by our model is consistent with the apparent experimental bound\cite{eisenstein-area} ($\gg
0.1$mm). This should be contrasted with the
vortex-free state which, as mentioned above, gives $\lambda_{J}\sim 5
\mum$ as the decay length.

The vortices in the emulsion will not be completely pinned, and hence
their presence will affect the counterflow superfluidity. Even if the
vortices remain pinned to the puddles they can move a distance $d_p$
across them, leading to a reduction in the stiffness.\cite{vavcrystals} Thermally activated hopping of vortices between
the puddles may lead to dissipation, as in previous work on the
coherence network,\cite{fertig,roostaei08} so that the emulsion may
formally be a vortex liquid at finite temperatures. However, the
distribution of $\ei$ in our model suggests a distribution of
activation energies, in contrast to previous work. 

Direct tests of our theory may be possible in imaging experiments.\cite{martin_localization_2004} For example, our model predicts that
charging lines corresponding to half-electron charges are common only
when $E_c\gtrsim E_s$. More generally, the identity of physical and
topological charge implies that the vortex configuration affects the
charging spectra.

Finally, let us revisit the role of layer-antisymmetric disorder. It
will give additional terms in Eq.~\ref{eq:esenergy} which are
proportional to $\sigma_i$, leading to terms linear in $v_i$ in the
Coulomb gas [Eq.~\ref{eq:disordermodel}]. Provided the compressible
puddles are effective at screening the antisymmetric disorder, the
energy of the charge imbalance $\sigma_i$ will be approximately
$e^2/C_M$, where $C_M \sim d_p^2/l_0$ is the mutual capacitance of two
puddles in opposite layers. This energy is a factor of $l_0/d_p \ll 1$
smaller than \ec, and the terms in $v_i$ are small compared with those
in $v_i^2$. Thus while layer-antisymmetric disorder could affect
correlations on very long scales, it will not affect the physical
consequences described above, which are controlled by the scale $d_p$.

\section{Conclusions}
\label{sec:conclusions}In conclusion, we have developed a model of a disordered quantum Hall
bilayer, in the experimentally relevant limit of strong, smooth
disorder. We have argued that the ground state of this model can be
understood as an emulsion of vortex-antivortex crystals. Our theory
suggests that the density of the emulsion could vary significantly
with magnetic length, and between samples, allowing its effects to be
isolated experimentally. An important physical consequence of the
presence of such an emulsion (or other disordered vortex state) is a
suppression of the decay of counterflow currents, potentially
explaining the area scaling of the tunneling
anomaly.\cite{eisenstein-area}

\vspace*{0.2in}This work was supported by EPSRC Grant No. EP/C546814/01. We thank
A. Stern for a critical reading of the manuscript, and P. Littlewood
for discussions.


\end{document}